\documentclass[10pt,aps,showpacs,twocolumn]{revtex4-2}
\usepackage{amsmath,amssymb,graphicx}
\usepackage{mathptmx}
\usepackage{color}

\usepackage{addfont}
\usepackage{lipsum}
\addfont{OT1}{rsfs10}{\rsfs}

\preprint{APS/123-QED}

\usepackage{xcolor}

\usepackage{textcomp}

\usepackage{lineno}

\usepackage[scr=rsfso,cal=zapfc,frak=euler,bb=ams]{mathalfa}

\begin{document}


\title{Calorimetry of photon gases in nonlinear multimode optical fibers}

\thanks{We acknowledge D.S.Kharenko, M. Gervaziev and Y. Sun for helping in the development of the mode decomposition setup.
The authors declare no conflicts of interest. }%

\author{M.~Ferraro$^{1,\dag}$}
\author{F.~Mangini$^{1,\dag,*}$}
\author{F.\,O.~Wu$^{2}$}
\author{M.~Zitelli$^{1}$}
\author{D.\,N.~Christodoulides$^{2}$}
\author{S.~Wabnitz$^{1,3}$}

\affiliation{$^1$ 
Department of Information Engineering, Electronics, and Telecommunications, Sapienza University of Rome, Via Eudossiana 18, 00184 Rome, Italy}
\affiliation{$^2$ CREOL - College of Optics and Photonics, University of Central Florida, Orlando, Florida 32816, USA}
\affiliation{$^3$ CNR-INO, Istituto Nazionale di Ottica, Via Campi Flegrei 34, 80078 Pozzuoli, Italy}
\affiliation{*Corresponding author: fabio.mangini@uniroma1.it}
\affiliation{$\dag$ These authors have contributed equally.}

\date{\today}
%
\begin{abstract}
\noindent
Because of their massless nature, photons do not interact in linear optical media. However, light beam propagation in nonlinear media permits to break this paradigm, and makes it possible to observe photon-photon interactions. Based on this principle, a beam of light propagating in a nonlinear multimode optical system can be described as a gas of interacting particles. As a consequence, the spatio-temporal evolution of this photon gas is expressed in terms of macroscopic thermodynamic variables, e.g., temperature and chemical potential. Moreover, the gas evolution is subject to experiencing typical thermodynamic phenomena, such as thermalization. The meaning of thermodynamic variables associated with the photon gas must not be confused with their classical counterparts, e.g., the gas temperature cannot be measured by means of standard thermometers. Although the thermodynamic parameters of a multimode photon gas result from a rigorous mathematical derivation, their physical meaning is still unclear. Specifically, one may question the significance of entropy, which lays at the basis of the thermodynamic theory. This is because no study has ever experimentally reported calorimetric processes of photon gases in multimode nonlinear optical systems. Entropy only acquires a physical meaning when heat exchanges between two systems are considered, because its maximization imposes that heat flows from hot to cold while it prohibits a converse scenario. 
In this work, we report on optical calorimetric measurements, which exploit nonlinear beam propagation in multimode optical fibers. Our results show that, indeed, heat only flows from a hot to a cold photon gas subsystem. This provides an unequivocal demonstration that nonlinear multimode wave propagation phenomena are governed by the second law of thermodynamics. In addition to be fundamental, our findings provide a new approach to light-by-light activated management of laser beams.
\end{abstract}
%
\maketitle
%
%
%
\section{Introduction}

Optical and thermodynamic phenomena are generally categorized as two distinct classes in classical physics, the former being ruled by the Maxwell equations and the latter following the principles of thermodynamics. Although thermodynamic effects merge with optics when light interacts with condensed matter, e.g., polariton condensates in semiconductors \cite{byrnes2014exciton} and plasmonic resonances in metallic nanostructures \cite{baffou2013thermo}, particle interactions which permit to reach a thermal equilibrium are virtually impossible in purely optical systems, due to the massless nature of photons. Nevertheless, in nonlinear optics, photons interact with each other via the nonlinearity of the propagation medium. In the linear regime, as a consequence of their superposition, electromagnetic waves interfere and, as soon as their superposition ends, the waves retrieve their initial state. On the other hand, in the nonlinear domain, electromagnetic waves do not simply interfere, but they may experience a nontrivial utterly complex interaction, so that their state will significantly differ from that at the input \cite{boyd2020nonlinear}. To understand  such complex interactions in multimode nonlinear systems, a series of attempts was made to establish a statistical description of these effects. Hence, owing to nonlinearity, a description of light propagation within a thermodynamic framework becomes possible \cite{onorato2015route, picozzi2014optical, guasoni2017incoherent, pierangeli2018observation, barviau2008toward, vanderhaegen2022observation, PhysRevLett.129.063901}.

In this context, the use of thermodynamics laws applied to nonlinear optical waves becomes particularly compelling when dealing with highly multimode systems, whose description in terms of statistical mechanics has been recently proposed in \cite{wu2019thermodynamic}. In this framework, a multimode nonlinear arrangement exhibits a behavior akin to that of a  gas of particles. In this respect  the photon occupancies assigned to each mode  varies during nonlinear propagation-an aspect that leads to energy exchange via particle collisions. Eventually, the mode occupancy distribution reaches equilibrium, which corresponds to the state of maximum entropy ($S$). This distribution can be described by the Rayleigh-Jeans (RJ) law, when the modes are sorted by their propagation constant. This process is associated with the definition of a temperature ($T$) and a chemical potential ($\mu$), that uniquely determine the state of thermal equilibrium. Such an equilibrium is described by an equation of state, which links $T$ and $\mu$ to the parameters of the system, i.e.: (i) the Hamiltonian, which is related to the values of the mode propagation constants and  plays the role of the internal energy ($U$); (ii) the optical power of the beam, which plays the role of the number of particles or, equivalently, of the mass ($\mathcal{P}$); (iii) the number of modes supported by the systems, which corresponds to the volume that is accessible to the photon gas ($M$):
\begin{equation}
    U-\mu \mathcal{P} = MT
    \label{eq-state-eq}
\end{equation}

This theoretical framework applies to any type of nonlinear multimode system (as long as this has a highly multimode nature and the nonlinearity is relatively weak). Recently, its validity has been experimentally confirmed in multimode optical fibers (MMFs) \cite{pourbeyram2022direct,mangini2022statistical}. When applied to the relevant case of graded-index MMFs, the thermodynamics of nonlinear beam propagation allows one to describe the so-called beam self-cleaning effect, whereby a highly speckled beam morphs into a bell-shaped wavefront because of nonlinearity \cite{krupa2017spatial,liu2016kerr}. Such an effect has attracted a great deal of interest in recent years, owing to its potential relevance for applications, e.g., for the development of high-energy fiber lasers \cite{wright2017spatiotemporal,tegin2020single} and high-resolution imaging devices \cite{moussa2021spatiotemporal}.

It is in this context that the thermodynamics of nonlinear MMFs shows its real potential, since it one to predict from the mere knowledge of the input beam state, what is ultimately going to be the mode occupancy at the output of a MMF, without recurring to a statistical analysis based on time- and energy-consuming computational tools. In this regard, it is important to note that the values of $T$ and $\mu$ are fully determined by the input coupling conditions between the laser source and the MMF. Therefore, in practical applications, the temperature and the chemical potential of a beam can be varied by simply introducing a small angle between the direction of the laser beam and the fiber axis \cite{podivilov2022thermalization}, as well as by modifying the laser wavefront by introducing an obstacle such as an aperture or a scattering medium \cite{baudin2020classical}. Generally speaking, the lower one sets the beam temperature, the “cleaner” the thermalized beam will look, i.e., the highest the output beam quality or brightness will be. If the temperature is low enough, only the fundamental mode, i.e., the mode having the highest propagation constant, is going to be macroscopically populated at the equilibrium, which gives rise to a phenomenon known as classical wave condensation \cite{connaughton2005condensation,aschieri2011condensation}. 

Here, we focus on demonstrating a crucial prediction of the statistical mechanics approach for describing nonlinear beam propagation in MMFs \cite{wu2019thermodynamic}. Specifically, we extend the theoretical study of thermodynamics to calorimetry, by considering the process of heat transfer between two photon gases which propagate in a MMF. Their interaction leads to the establishment of a new state of thermal equilibrium, whose parameters (temperature and chemical potential) are fully determined by the initial conditions of the two gases, i.e., by the temperature and chemical potential of each gas separately. We show that theoretical predictions are fully confirmed by experiments, based on a holographic mode decomposition technique \cite{gervaziev2020mode}. These results extend the potential of the beam self-cleaning effect in terms of applications. Indeed, our findings imply that is possible to modify the quality or brightness of a laser beam, by exploiting its nonlinear interaction in a MMF with another beam with either higher or lower temperature. In this sense, this result paves the way for the development of a novel generation of photonics tools for the all-optically controlling the spatial profile of intense laser beams.

\section{Theoretical framework}
\begin{figure*}[htb]
  \centering
  \includegraphics[width=16.7cm]{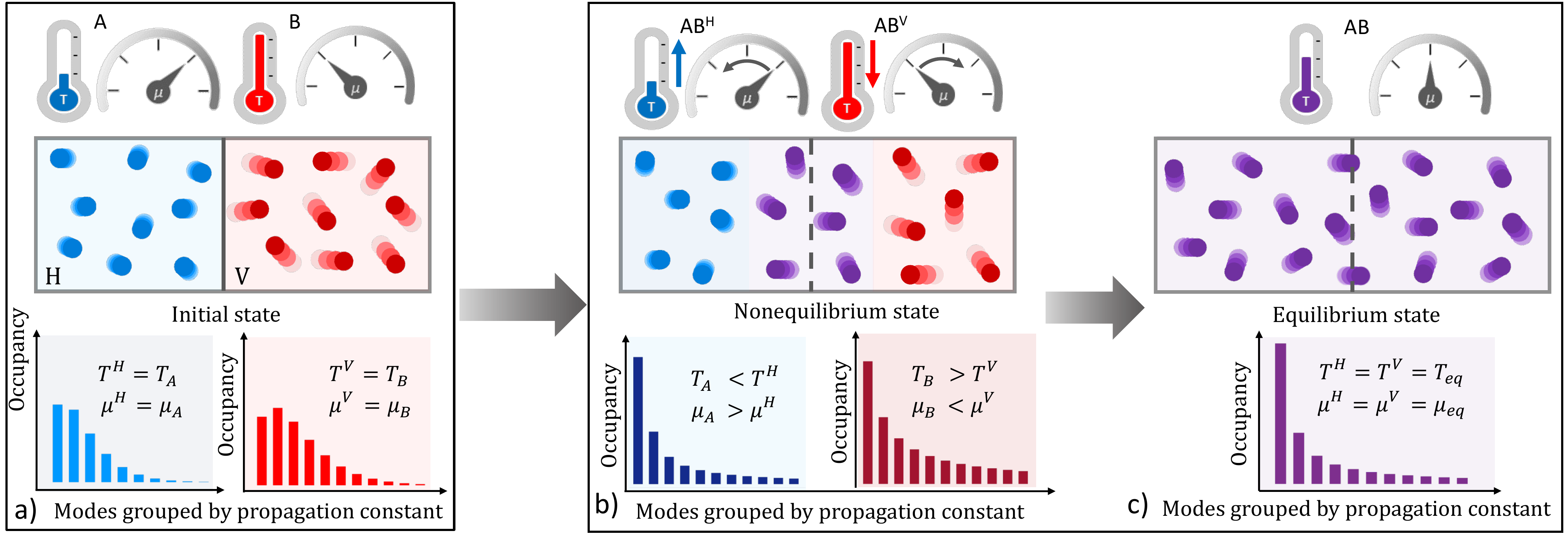}
\caption{Sketch of the thermalization process of two photon gases. (a) Initial state: two orthogonally polarized beams are prepared with different temperatures, chemical potentials, and mode distributions. (b) Nonequilibrium transient state: the difference in temperature and chemical potential among the left and right sides of the box progressively quenches, until their associated mode occupancy progressively approaches the RJ distribution, as a consequence of nonlinear beam propagation, i.e., of heat exchange between the two gases. (c) Equilibrium state: the two gases reach the same values of temperature and chemical potential, and their mode distribution obeys the RJ law.}
  \label{fig-toy}
\end{figure*}

Thermodynamics rules the properties of gases at thermal equilibrium, and permits their description in terms of macroscopic parameters such as $M$, $\mathcal{P}$, $\mu$, and $T$ \cite{wu2019thermodynamic,fermi2012thermodynamics}. Whereas, the study of modifications of thermal equilibrium by, e.g., the exposition of a given object to an external heat source, or via a phase transition, is the key topic of calorimetry. The definition of the temperature of a gas has, on the one hand, a pure mathematical meaning in thermodynamics. In fact, it is associated with the statistics that describes the displacement of microscopic particles inside a volume, or, in the case of a photon gas, their mode occupancy. On the other hand, when placed in the context of calorimetry, temperature acquires a physical connotation. Indeed, when nonequilibrium states are considered, temperature becomes the physical parameter that varies, owing to the exchange of heat (i.e., of energy) between two objects. Specifically, heat can only flow from an object with higher temperature towards the object with lower temperature \cite{fermi2012thermodynamics}. The heat flow will only stop when a new equilibrium is established, i.e., when both objects have reached the same temperature ($T_{eq}$). 

It is important to underline that such a principle is associated with the maximization of entropy $S$, i.e., with the second principle of thermodynamics, whereas the equilibrium statistics of a gas in thermodynamics is only determined by imposing the stationarity of $S$. Therefore, experimentally demonstrating the establishment of an equilibrium as a consequence of the interaction (heat exchange) between two photon gases is a key for proving that the thermodynamic temperature associated with the RJ mode distribution has indeed a true physical meaning.

In this work, we demonstrate that by considering the simplest case of two optical beams (labelled as A and B), which are prepared with different temperatures ($T_A$ and $T_B$) and chemical potentials ($\mu_A$ and $\mu_B$), simultaneously propagating in an MMF. These results are accord in previous theoretical predictions where the internal energy $U$ and power $P$ is allowed to resettle between two subsystems \cite{wu2019thermodynamic}. Here, we prepared the two beams with orthogonal vertical and horizontal linear states of polarization, respectively. In this way, the initial state of the system can be depicted as that of two gases which occupy two complementary portions of the total available mode volume (see Fig. \ref{fig-toy}a): these correspond to horizontally and vertically polarized modes. In each side of the box, the color of the beams indicates their temperature. Specifically the colder gas, depicted in blue, corresponds to the horizontally polarized beam (A), whereas the hotter gas, depicted by red particles, is associated with the vertically polarized beam (B). We note that the use of colors is meant to help the visualization of the phenomenon. As a matter of fact, the gases are made of photons which, being bosons, are fully indistinguishable.

Moreover, it has to be noted that under sufficiently adiabatic conditions it is possible to associate a photon gas with a temperature even when the gas is in a nonequilibrium state, i.e., its associated mode occupancy distribution does not yet follow the RJ law, as depicted at the bottom of Fig. \ref{fig-toy}a. Indeed, one can compute the value of the temperature on each side of the box in Fig. \ref{fig-toy}a from its corresponding mode occupancy distribution, as described in \cite{wu2019thermodynamic}. A laser beam propagating in the MMF core is isolated, since the energy losses into the cladding are negligible in the absence of fiber bending \cite{gloge1972bending}. As a consequence, in the presence of a single beam, the nonlinear beam dynamics in an MMF can be described as the free expansion of a gas. This aspect will be of interest in the next Sections.

We will use the superscript letters 'H' and 'V' when referring to the parameters of the left and right side of the box, respectively. On the other hand, when referring to the initial temperature of each gas, we will use as subscript the letters 'A' and 'B', respectively. Therefore, the initial state in Fig. \ref{fig-toy}a is defined by $T^H = T_A$, $\mu^H = \mu_A$, $T^V = T_B$, and $\mu^V = \mu_B$.

Once the photon system evolves, the two gases interact, until progressively reaching the same equilibrium parameters. In our picture, the system evolution is enabled by the removal of the barrier between the left and right sides of the box. In particular, it must be noted that removing the barrier allows for the mixing of particles which originally belong to either the A or the B photon gases. Therefore, the thermalization process must be described in terms of the grand canonical ensemble. The chemical potential of each gas will vary: eventually, as it occurs for the temperature, the chemical potential of the two gases will be the same ($\mu_{eq}$). 
The other way around, if exchange of particles between A and B gases is not allowed, e.g., by means of a barrier which is only permeable to heat, the two gases would remain distinguishable during the whole thermalization process, which would have needed to be described in terms of the canonical statistical ensemble. In the canonical ensemble regime, each gas subsystem would conserve its number of particles during the heat exchange process, and $\mu^H$ and $\mu^V$ do not converge to the same value at the occurrence of thermalization.
As we will see in the following, in this work we consider the former case in our experiments. However, we will provide scenarios for possible experimental demonstrations of the beam heating and cooling in the framework of a canonical ensemble.

It is reasonable to assume that the equilibrium value of the temperature and chemical potential (which are associated with violet particles in Figs. \ref{fig-toy}b-c) is first reached in the center of the box, where the first interactions between the two gases take place (see Fig. \ref{fig-toy}b). Later on, a thermal equilibrium is reached at the box periphery (see Fig. \ref{fig-toy}c). In its nonequilibrium transient state, each side of the box is associated with a mode distribution which progressively approaches the RJ law (see the histograms at the bottom of Figs. \ref{fig-toy}b,c). Of course, as the color map shows, before reaching an equilibrium the temperatures on the two sides of the box are different. Indeed, in the left part of the box the coldest gas is still present. As a result, the temperature of the horizontally polarized photon gas will be higher than that of the photons in vertically polarized modes, because its associated gas still contains hotter particles (i.e., $T_A<T^H \leq T^V<T_B$, and $\mu_A>\mu^H \geq \mu^V>\mu_B$). The temperature difference between the left and right sides of the box progressively quenches as the input power grows larger, and eventually vanishes when a thermal equilibrium is reached (i.e., $T^H=T^V=T_{eq}$ and $\mu^H=\mu^V=\mu_{eq}$).

The values of $T_{eq}$ and $\mu_{eq}$ can be easily calculated. As a matter of fact, when a gas of mass $\mathcal{P}$ increases its temperature from $T>0$ to $T+dT$, it absorbs a quantity of heat $\delta Q$ which is given by
\begin{equation}
    \delta Q = \mathcal{P} \cdot c  \cdot dT,
    \label{eq-delta-q}
\end{equation}
where c is the heat capacity of each gas, which is supposed to be independent of $T$ \cite{wu2020entropic}. In the case of two gasses interacting in an isolated system, we can impose that $Q_A = - Q_B$, i.e., all of the heat absorbed by A is provided by B, which leads to
\begin{equation}
T_{eq}  =  \frac{\mathcal{P}_A  T_A  + \mathcal{P}_B  T_B}{\mathcal{P}_A  + \mathcal{P}_B},	
\label{eq-T-eq}
\end{equation}
where we have reasonably assumed that the two gases have the same heat capacity. Note that Eq. (\ref{eq-T-eq}) holds as long as both $T_A>0$ and $T_B>0$. As a matter of fact, in the thermodynamic description of multimode optical systems, thermal equilibrium can be reached even at negative temperatures \cite{wu2019thermodynamic}. It is convenient to define the relative power of the B beam as $R = \mathcal{P}_B  / (\mathcal{P}_A+\mathcal{P}_B)$, so that the equilibrium temperature more easily reads as $T_{eq}  = (1-R)  T_A  + R T_B$. Finally, the value of $\mu_{eq}$ can be easily determined by substituting expression (\ref{eq-T-eq}) in the equation of state (\ref{eq-state-eq}).

\section{Experimental methods}

Short spans of a graded-index MMF turn out to be the perfect testbed for the verification of the process of heat exchange between two photon gases. As a matter of fact, at near-infrared wavelengths, MMFs are an adiabatic system since their optical losses are practically negligible, as long as their length is of the order of a few meters \cite{agrawal2000nonlinear}, and the beam optical power is far from the catastrophic self-focusing regime \cite{ferraro2021femtosecond}. As such, MMFs have been employed for the demonstration of classical wave thermalization \cite{pourbeyram2022direct,mangini2022statistical}. 

Here, we use a 2 m long span of a standard graded-index MMF (GIF50E from Thorlabs), whose core diameter is 50 $\mu$m. Such a fiber supports about 110 modes at the wavelength of our laser (1030 nm, Pharos from LightConversion, emitting 2 ps pulses with peak powers of the order of tens of kW). 

Two laser beams with a transverse diameter of 30 $\mu$m are simultaneously injected into the fiber core. The two beams form different small angles (less than 2${}^\circ$) with respect to the fiber axis, in order to be associated with different temperatures. Both beams are injected at the core center, in order to avoid the seeding of a nonzero orbital angular momentum, which would be responsible for a modification of the equilibrium distribution \cite{wu2022thermalization}. In addition, the temperature of beam B is varied by introducing an aperture along the laser path (see Supplementary Note 1). The two beams are orthogonally polarized, so that any interference at the fiber input facet is avoided: the total power injected into the fiber is simply given by the sum of powers in each of the two beams ($\mathcal{P}_{AB}= \mathcal{P}_A+\mathcal{P}_B$). 

At the fiber output, we employed the mode holographic decomposition tool as described in \cite{gervaziev2020mode}. This allowed for retrieving the values of the mode occupancy at the fiber output, which permits to determine the temperature and chemical potential associated with each photon gas \cite{wu2019thermodynamic}. 

\section{The simple case of free expansion of a photon gas}

\begin{figure*}[htbp]
  \centering
  \includegraphics[width=14cm]{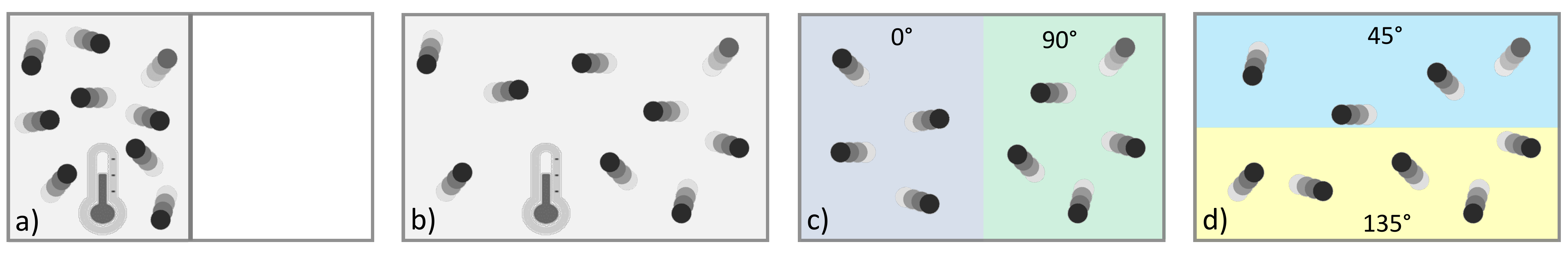}
  \includegraphics[width=15.5cm]{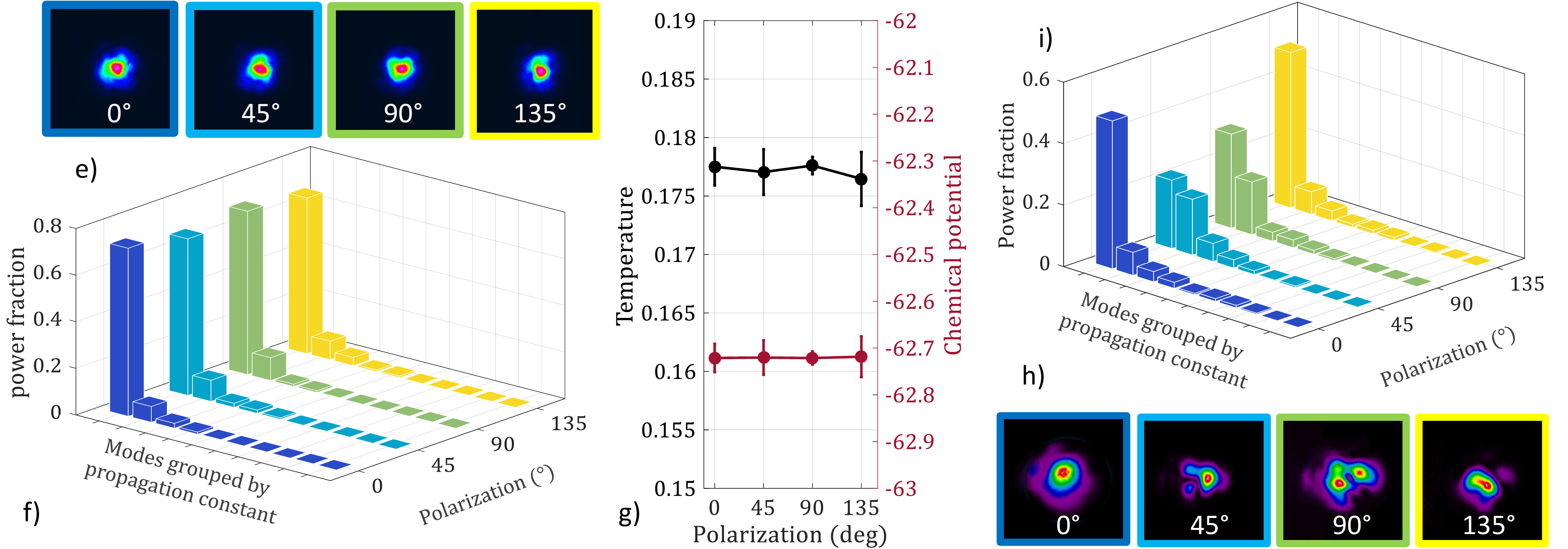}
\caption{Free expansion of a photon gas. a) Input condition associated with a horizontally polarized laser beam. b) Result of the gas free expansion. c,d) Output polarization analysis corresponding to either horizontal/vertical (c) or 45${}^\circ$/135${}^\circ$ (d) polarization splitting. e,f) Output beam profiles (e) and their associated normalized mode occupancy (mode power fraction) (f) at input peak power of 14 kW along different output polarization directions. g) Temperature and chemical potential associated to the RJ distributions in (f). h,i) Same as (e,f) at the input peak power of 7 kW.}
  \label{fig-free-exp}
\end{figure*}

Let us first consider the simplest case when just a single photon gas is present, that freely expands into a box while doubling its volume (Figs. \ref{fig-free-exp}a,b). This picture corresponds to the injection of a beam which has a horizontal linear polarization at the fiber input. In this case, initially the gas only occupies half of the whole modal volume, whereas the other half remains empty (Fig. \ref{fig-free-exp}a). As a consequence of its nonlinear propagation into the MMF, the beam eventually loses its linear degree of polarization at the fiber output: modes with vertical polarization also get populated, and the gas uniformly occupies the whole box volume (cfr. Fig. \ref{fig-free-exp}b).

The free expansion of the photon gas can be experimentally demonstrated by projecting the beam at the fiber output onto a given polarization direction by means of a polarizing beam splitter. 
In particular, here we illustrate two significant cases, i.e., corresponding the insertion of either a vertical or a horizontal barrier in the box. In analogy with Fig. \ref{fig-free-exp}a, a vertical barrier separates the mode volume into horizontally and vertically polarized modes, which are highlighted in Fig. \ref{fig-free-exp}c by grey and light green colors, respectively. Whereas the horizontal barrier projects the modes onto the 45${}^\circ$ and the 135${}^\circ$ polarization directions, as highlighted in Fig. \ref{fig-free-exp}d by light blue and yellow colors, respectively. 

The intensity profile of the output beam for the four polarization directions is reported in Fig. \ref{fig-free-exp}e. Specifically, these results were obtained when injecting laser pulses of 14 kW of peak power. As it can be seen, a bell-shape (which is associated with a state of thermal equilibrium) is found along all polarization directions, owing to the occurrence of the beam self-cleaning effect. This is in agreement with former studies of the nonlinear polarization dynamics of beam self-cleaning \cite{krupa2019nonlinear}. For the sake of completeness, in Fig. \ref{fig-free-exp}f we also show the mode distributions associated with the images in Fig. \ref{fig-free-exp}e. One cannot help noticing that the differences between all of these mode distributions are practically negligible. Thus, they are associated with virtually identical values of $T$ and $\mu$. This is can be appreciated from the data in Fig. \ref{fig-free-exp}g, where we report the values of temperature and chemical potential which are calculated by starting from the experimental mode decomposition for different projection directions of the output state of polarization. Indeed, the temperature and the chemical potential maintain constant values, within the experimental error.

It is worth noticing that, as mentioned above, one may formally calculate the temperature associated with a speckled beam from its associated mode occupancy values $U$, $P$, even if the beam has not yet reached thermal equilibrium. As a matter of fact, the thermodynamic parameters are fully defined by the coupling conditions of the laser beam at the fiber input, and do not vary as a consequence of the nonlinear propagation of the beam.
This is confirmed by the results in Figs. \ref{fig-free-exp}h,i. These correspond to a value of the input peak power of 7 kW, which is not high-enough for achieving beam self-cleaning along all states of polarization. Indeed, as it can be seen in Fig. \ref{fig-free-exp}h, rather different intensity profiles of the output beam are obtained for the four polarization directions. Specifically, the beam only has a bell-shape in the horizontal state of polarization. Whereas, a speckled pattern is found when selecting other polarization directions. As a result, different mode occupancy distributions are found for each output intensity pattern (see Fig. \ref{fig-free-exp}g). However, the thermodynamic parameters corresponding to all of these distributions turn out to be exactly the same, in agreement with theoretical expectations. 

Finally, we underline that, in our representation the role of time is played by the fiber length, which is fixed in the experiment. Whereas, the beam power rules the “speed” of energy exchange among particles, which belong to either the same gas or to different gases. Therefore, we may track the “temporal” evolution of the temperature of the box as a function of input power. It must be underlined that, as mentioned above, the input power also defines the number of particles in the gases. Therefore, in order to emphasize the role of input power in the interaction between the gases, in the following we will refer to the temperature as the ratio between $T$ and $\mathcal{P}$. In this regard, one may notice that this operation does not invalidate the theoretical expectation value of the equilibrium temperature, which only depends on the power ratio between A and B (see Eq. (\ref{eq-T-eq})), whose value is fixed all of our experiments.

\section{Reaching thermal equilibrium when mixing two photon gases with the same ``mass"}

In this section, we consider the thermalization process of two beams carrying the same power (i.e., $R$ = 1/2). Correspondingly, the two gases are formed by the same number of particles, i.e., they have the same ``mass".  
Aiming at emphasizing the role of heat exchange between the two gases, we compare three different injection conditions. Specifically, we consider the injection of the sole A beam, the injection of the sole B beam and the injection of both. In order to distinguish the corresponding thermodynamic parameters, in the following, these three cases will be associated with the subscript symbols 'A', 'B', and 'AB', respectively (e.g., the temperature of the left side of the box when only the A beam is present will be referred to as $T^H_A$).

\begin{figure}[htbp]
  \centering
  \includegraphics[width=8.65cm]{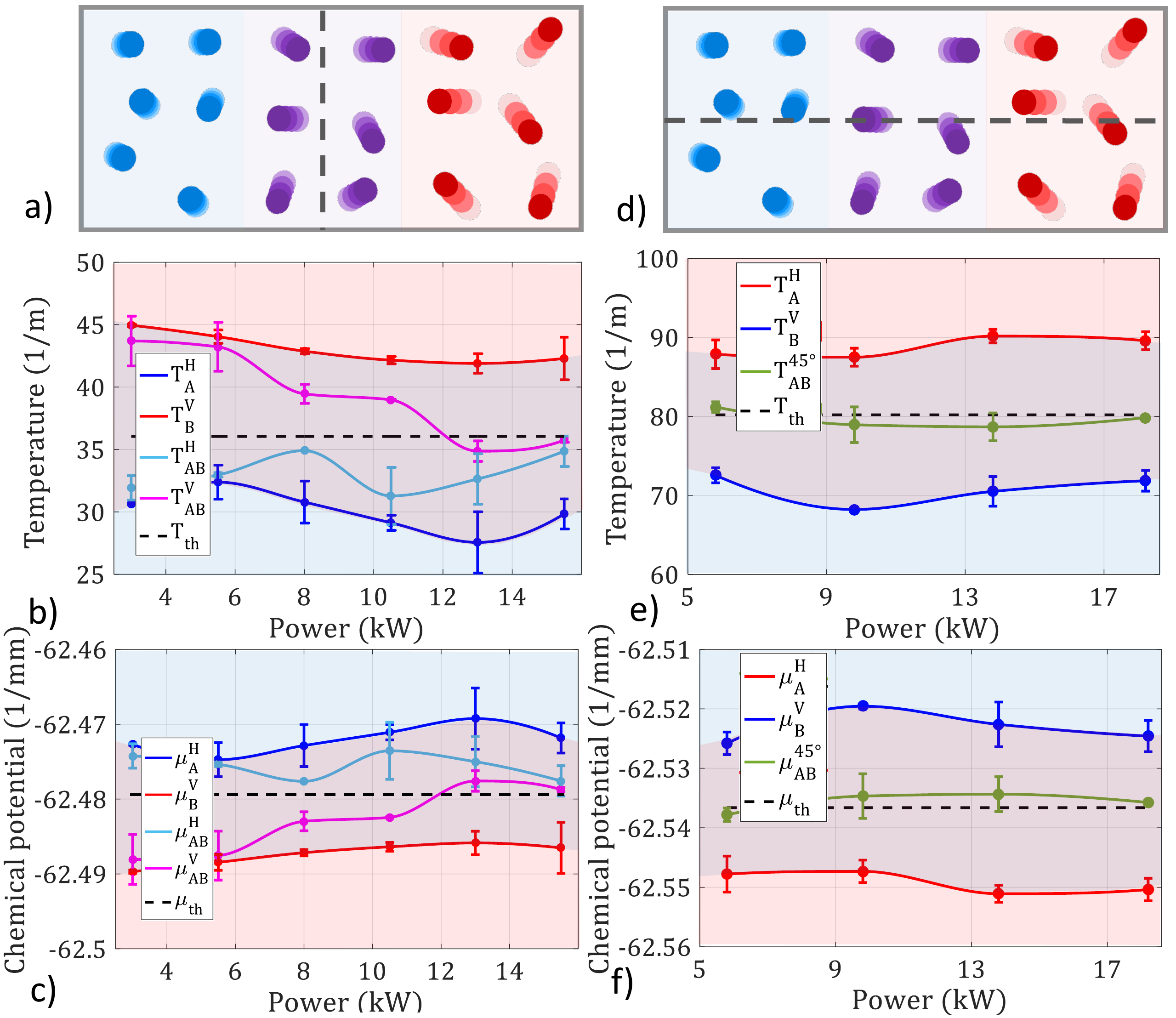}
\caption{Thermal equilibrium in the presence of two photon gases with the same mass ($R$ = 1/2). a-c) Schematic representation of polarization analysis corresponding to horizontal/vertical polarization splitting (a); corresponding measured values of temperature (b) and chemical potential (c), either in presence of the sole A and B beams, or when simultaneously injecting A and B beams into the MMF core. d-f) Same as (a-c) for 45${}^\circ$/135${}^\circ$ polarization splitting.}
  \label{fig-res}
\end{figure}

The experimental results in this case are shown in Fig. \ref{fig-res}. At first, we inserted a horizontal/vertical beam splitter at the fiber output, which is analogous to inserting a vertical barrier into the box, as depicted in Fig. \ref{fig-res}a. The corresponding values of the temperature and the chemical potential at both sides of the box are shown in Fig. \ref{fig-res}b,c (see Supplementary Note 2 for details about the calculation of thermodynamic parameters). In agreement with theoretical expectations, we found that, in the presence of both beams, the temperature at the left side of the box is always higher than the temperature measured in the presence of the sole A beam, for all values of the input power, i.e., $T^H_A \geq T^H_{AB}$
. The other way around, symmetrically, we have that $T^V_B \geq T^V_{AB}$ ($\mu^V_B \leq \mu^V_{AB}$).
In particular, at low input powers, i.e., in the quasi-linear regime, the temperature at the left side of the box remains the same, within the experimental error, independently on whether we inject the sole A beam, or both A and B beams ($T^H_A \simeq T^H_{AB}$). 
To the contrary, as the input power grows larger, the $T^H_{AB}$ curve progressively drops down, until it reaches a plateau value, which coincides with that of $T^V_{AB}$, i.e., the temperature at the right side of the box, which is obtained when both A and B beams co-propagate into the MMF. This indicates that an equilibrium temperature is reached, whose value is in good agreement with the theoretical expectation of Eq. (\ref{eq-T-eq}) (dashed black line in Fig. \ref{fig-res}b). 

It is worth noting that the temperature at both sides of the box does not vary with input power in the case where only one beam is present, i.e., both the $T^H_A$ and $T^V_B$ curves are flat in Fig. \ref{fig-res}b. This confirms that the nonlinear beam dynamics in a MMF can indeed be described as the free expansion of a gas, as discussed in the previous Section. 

The evolution of the chemical potential follows the same trend of the gas temperature. In Fig. \ref{fig-res}c, we can see that $\mu^H_{AB}$ and $\mu^V_{AB}$ both reach the same value at high input powers. Whereas at low input powers their values are close to that of $\mu^H_A$ and $\mu^V_B$, respectively. This further proves that a thermalization phenomenon between two indistinguishable gases is observed. 
In this regard, it is worth mentioning that distinguishing the beams through their polarization state was the easiest solution that could be adopted with our experimental setup. Nevertheless, the theoretical principle behind beam cooling and heating goes beyond this specific case. For example, one may think of using two beams emitted by laser sources that differ in their wavelength or pulse duration, thus enabling the observation of thermalization within the canonical ensemble.

In order to further confirm the validity of the representation in Fig. \ref{fig-toy}, we analyzed the
output beam projections along the 45${}^\circ$/135${}^\circ$ polarization directions. Indeed, if the representations in Figs. \ref{fig-res}a is valid, one would expect to observe a temperature in the upper and lower sides of the box which does not depend on input power. Indeed, the average temperature of the top and bottom sides of the box in Fig. \ref{fig-res}d is expected to be the same, since both sides contain the same amount of hot, cold, and thermalized particles, although the system is clearly not yet at the equilibrium. This is in agreement with the experimental results reported in Fig. \ref{fig-res}e,f, where we show that both temperature and chemical potential obtained when simultaneously injecting A and B beams maintain nearly constant values as the input power is varied, in agreement with theoretical predictions. Such a behavior is the opposite of what is observed in Fig. \ref{fig-res}b,c, thus validating our model of gas thermalization, which is illustrated in Fig. \ref{fig-toy}. As a final note, we highlight that the values of $T$ and $\mu$ in Figs. \ref{fig-res}b-c and Figs. \ref{fig-res}e-f are different, simply because the experimental results reported therein were carried out in different days, i.e., with different experimental input laser-fiber coupling conditions.

\section{Entropy growth}
One may notice that in the pictures in Fig. \ref{fig-res}a we are neglecting the spreading of the red gas into the right side of the box (as depicted in Figs. \ref{fig-free-exp}a,b). In this regard, it must be emphasized that the results in Figs. \ref{fig-res}b,c demonstrate that, even if present, the spreading of the gases into opposite sides of the box is a slower process when compared with the change of temperature of their particles. In other words, the energy exchange between modes is a more efficient process than nonlinear depolarization. 
Moreover, the possible spreading of non-thermalized particles on the opposite side of the box, e.g., that given by linear depolarization, would not invalidate the theory of photon beam calorimetry. Indeed, the establishment of the equilibrium is not only determined by the measurement of one and the same temperature. As a matter of fact, thermal equilibrium is properly reached if and only if the mode occupancy distribution on the left and right sides of the box is identical, and follows the RJ law, i.e., only if the entropy has reached a maximum. This result has been experimentally verified (see Supplementary Note 3). Indeed, we found that only at the highest value of the input power (i.e., 16 kW) the distribution associated with both horizontally and vertically polarized modes properly obeys the RJ law. 

\begin{figure}[htbp]
  \centering
      \includegraphics[width=8.6cm]{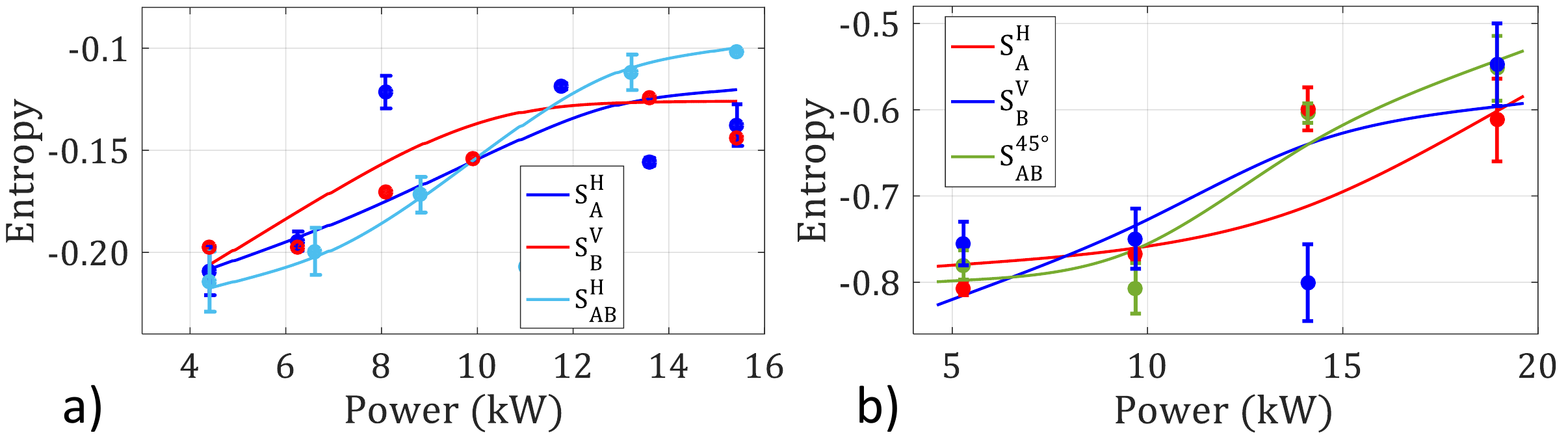}
\caption{Experimental demonstration of the entropy growth. a,b) Values of the entropy calculated from the data in Fig. \ref{fig-res}b,c and \ref{fig-res}e,f, respectively. The curves are simply guides for the eye.}
  \label{fig-entropy}
\end{figure}

At the same time, we verified the crucial condition that entropy grows larger when the input power grows lager. In particular, in Figs. \ref{fig-entropy}a,b we illustrate the measured variation of entropy with input laser power, corresponding to experimental results in Figs. \ref{fig-res}b,c and \ref{fig-res}e,f, respectively.
Following the methods of Ref. \cite{wu2019thermodynamic}, the entropy can be calculated as
\begin{equation}
    S = \sum_{i=1}^M \ln n_i,
\end{equation}
where the index $i$ runs over all groups of modes having the same propagation constant, whose occupancy is indicated as $n_i$. Further details on the method for computing the entropy are reported in the Supplementary Note 2.

As it can be seen in Figs. \ref{fig-entropy}a,b in all cases the entropy clearly experiences a growth when moving from low to high input powers, i.e., from an out-of-equilibrium to a state of full thermal equilibrium. In particular, the green curve in Fig. \ref{fig-entropy}b indicates that, although its temperature is equal to $T_{eq}$ (cfr. Fig. \ref{fig-res}e), thermal equilibrium in the top side of the box in Fig. \ref{fig-res}d is not reached at low powers, since its associated entropy is not maximized yet. This further confirms that, even if present, effects responsible for the diffusion of a gas into the whole modal volume, i.e., linear or nonlinear depolarization, cannot force the system to reach thermal equilibrium: this can only be reached thanks to heat exchange between the two gases.

\section{How to tune the equilibrium temperature}

\begin{figure}[htbp]
  \centering
  \includegraphics[width=8.65cm]{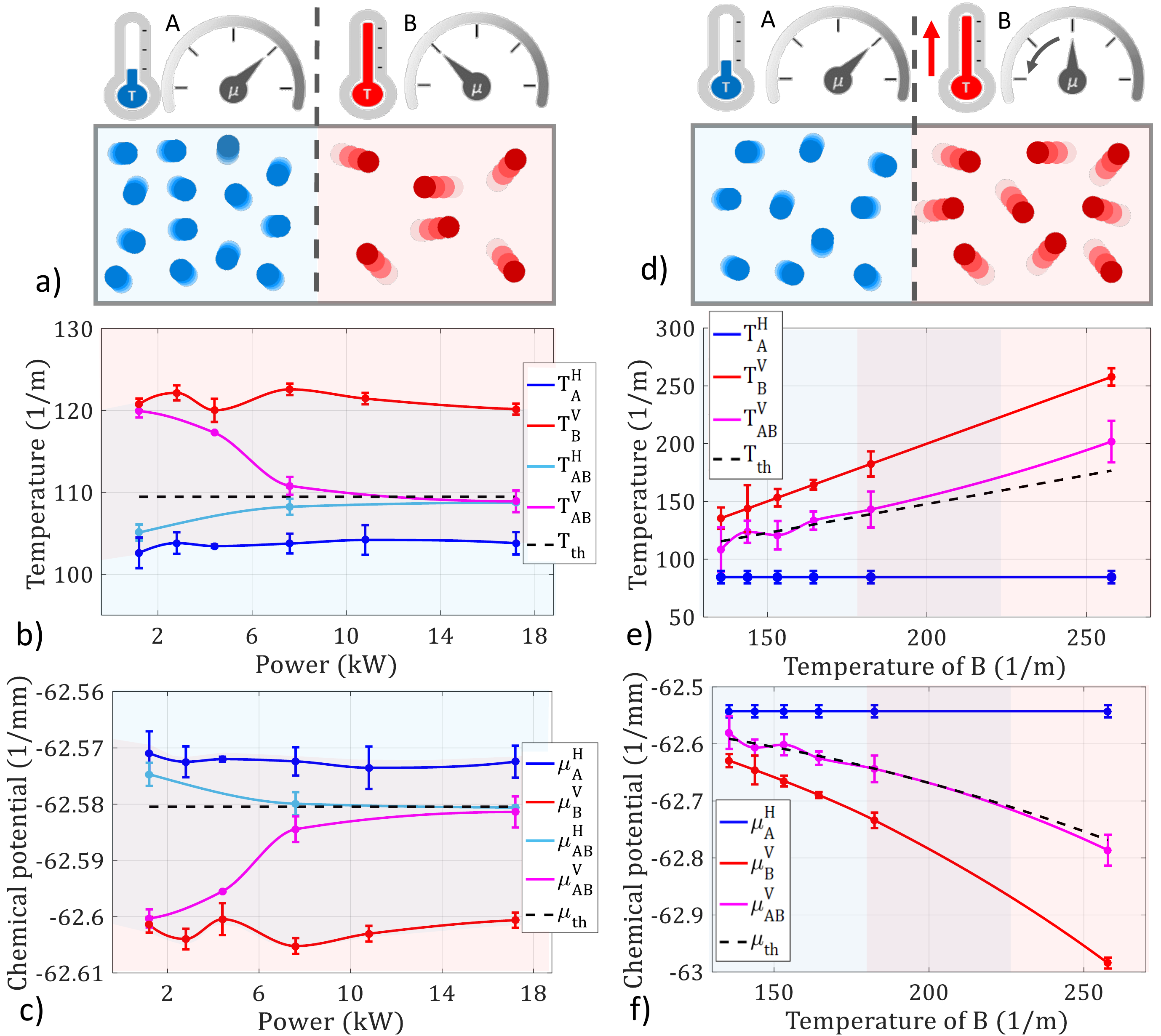}
\caption{Tuning of the equilibrium temperature of interacting photon gases. a-c) Same as Fig. \ref{fig-res}, when setting $R$ = 1/3. d) Sketch of the input state when varying  $T_B$ with respect to Fig. \ref{fig-res}a- e,f) Variation of the temperatures (e) and chemical potentials (f) at the left and right sides of the box, vs. $T_B$. In order to ensure system thermalization, in (e-f) the input power is equal to 16 kW, and $R$ = 1/2.}
  \label{fig-iris}
\end{figure}

Finally, let us demonstrate how it is possible to control the equilibrium temperature of the gas mixture. By recalling Eq. (\ref{eq-T-eq}), it appears evident that there are only two ways of tuning the equilibrium temperature, i.e., by acting on either $R$, or on the initial temperature of the two gases. Here, we experimentally demonstrate both these possibilities, as illustrated in Figs. \ref{fig-iris}a-c and \ref{fig-iris}d-f, respectively. 
In Fig. \ref{fig-iris}b,c, we report the results of the same experiment that was shown in Fig. \ref{fig-res}b,c, but now setting $R$ = 1/3 . As it can be seen, even in this case there is a good agreement between the experimental results and the theoretical expectations (black dashed curve). The equilibrium temperature is closer to that of the gas with the higher mass (B), as it occurs for gases made of matter particles. 
Finally, we varied the temperature of the beam B ($T_B$), while keeping constant its associated number of particles. In Fig. \ref{fig-iris}e we show that the equilibrium temperature linearly scales with $T_B$, as expected from theory. In particular, in the experiment reported in Fig. \ref{fig-iris}e, we prepare the beams so that $R$ = 1/2: here the input power is high enough to ensure full gas thermalization (as in Fig. \ref{fig-toy}c), except for the point corresponding to the highest value of $T_B$. Clearly, the theoretical value of the thermal equilibrium temperature of the gas mixture is always found at the mid-point between $T_A$ and $T_B$. Analogously, the chemical potential of the mixture is the average of the chemical potentials of the A and B beams (cfr. Fig. \ref{fig-iris}f).

\section{Conclusion}
We extended the thermodynamic theory of nonlinear beam propagation in optical multimode fibers by employing calorimetric methods, thus demonstrating the possibility of optical beam cooling and heating. This verifies that the thermodynamic parameters $T$ and $\mu$ not only have a rigorous mathematical definition, but also represent meaningful physical parameters. Indeed, our experimental results prove that the statistical mechanics approach to optical multimode systems respects the thermodynamics principles of energy conservation and entropy growth. In this work, we studied the heat transfer between two beams with different states of polarization, in the framework of the grandcanonical ensemble. This paves the way for the demonstration of calorimetry in the canonical ensemble, e.g., by studying the interaction of two beams with different wavelengths. In terms of applications, this will lead to the possibility of demonstrating laser beam quality improvement at a given wavelength by only acting on a second beam at a different wavelength, a property that would be of great interest for all-optical beam control applications, such as high-power fiber lasers and spatial division multiplexed optical communications. \\ \\

\begin{acknowledgments}
\noindent
MF, FM, MZ, and SW acknowledge the European Research Council (ERC) under the European Union’s H2020 and HORIZON EUROPE research and innovation programs (740355, 101081871) and Sapienza University (AR2221815ED243A0, AR2221815C68DEBB). The work of DNC and FOW has been partially supported by the Office of Naval Research (ONR) (MURI: N00014-20-1-2789), National Science Foundation (NSF) (EECS-1711230, DMR-1420620, Qatar National Research Fund (QNRF) (NPRP13S0121-200126), MPS Simons collaboration (Simons Grant No. 733682), W. M. Keck Foundation, US-Israel Binational Science Foundation (BSF: 2016381).
\end{acknowledgments}

\bibliography{Biblio}
\newpage
\noindent{\Huge \textbf{Supplementary \\ Notes}}\\
\section{Supplementary Note 1: \\Preparation of the gas initial state}
\begin{figure}[ht!]
\centering\includegraphics[width=9cm]{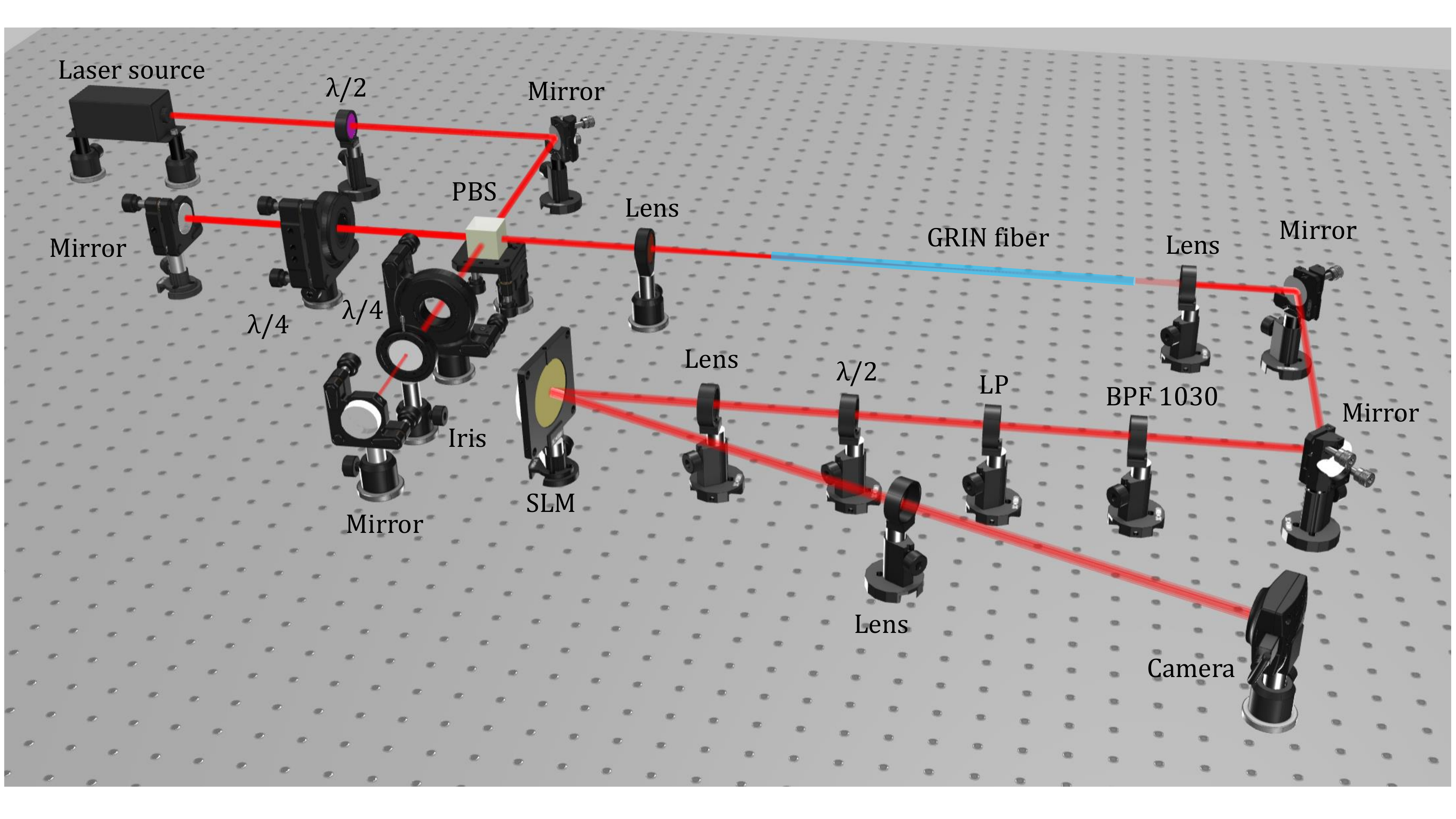}
\caption{Sketch of the experimental setup. $\lambda/2$: Half-waveplate; $\lambda/4$: Quarter-waveplate; 
PBS: Polarizing Beam Splitter; BPF 1030: Band Pass Filter at 1030 nm; LP: Linear Polarizer; SLM: Spatial Light Modulator.}

\label{fig-setup}
\end{figure}

Our experimental setup is shown in Fig. \ref{fig-setup}. Experiments were carried out by means of a Yb-based laser system (Light Conversion PHAROS-SP-HP), generating pulses of 2 ps with 100 kHz repetition rate, at 1030 nm. 
By means of a polarizing beam splitter, we separate the laser beam into an interferometer-like path, where each arm corresponds to a photon gas, as depicted in Fig. 1 of the main text. The setup was built in order to ensure that, once recombined, the two beams are temporally synchronized and with orthogonal linear states of polarization. The mirrors of one of the two arms was slightly tilted. In this way, we could prepare the two beams with different photon gas temperatures. The tilting angle was about 2${}^\circ$. We verified that introducing such an angle did not alter the transmitted power, i.e., the injection efficiency remained virtually the same. However, the intensity profile (at low powers) at the output of the fiber, as well as its associated value of temperature, could be significantly modified.

The power ratio $R$ between the two beams at the fiber input was determined by the orientation of a half-waveplate, which was placed upstream of the beam splitter. Whereas the variation of $T_B$, that we describe in Figs. 5 d-f in the main text, was possible by means of an iris placed in the arm B (see Fig. \ref{fig-setup}). The iris distorts the wavefront of the B beam, thus varying its associated temperature. Indeed, acting on the wavefront shape permits to modify the mode distribution (and, consequently, the thermodynamic parameters) associated with the initial state. However, the iris is also responsible for a loss of beam power, i.e., for the reduction of the mass of gas B. In order to ensure that the value of $R$ remained the same in the experiments of Fig. 5 of the main text, we compensated the power lost because of the iris by means of a quarter-waveplate, inserted in cascade with the iris (see Fig. \ref{fig-setup}). Note that another quarter-wavaplate was placed in arm A. This is because, in the absence of the quarter wave-plates, no light would reach the input tip of the MMF, since both arms would maintain their state of polarization, thus being reflected back towards the laser source by the polarizing beam splitter.

At the fiber output, the beam was collected by means of a lens, and analysed by means of a mode decomposition (MD) setup. This consists of a bandpass filter, a polarizer (or, equivalently, a polarizing beam splitter), a half-waveplate, two lenses, a spatial-light-modulator (Hamamatsu LCOS- X15213), and a CCD camera (Gentec Beamage-4M-IR). The working mechanism of the mode decomposition setup has been discussed in full details elseqhere, see for example Ref. \cite{gervaziev2020mode}.

\section{Supplementary Note 2: \\Calculation of the thermodynamic parameters from the experimental data}
The Boltzmann entropy is defined as
\begin{equation}
    S = \ln W,
    \label{eq-S}
\end{equation}
where W for bosonic systems can be written as \cite{wu2019thermodynamic}
\begin{equation}
    W(\{n_i\}) = \prod_{i=1}^M \frac{(n_i+g_i-1)!}{n_i!(g_i-1)!}.
    \label{eq-W}
\end{equation}
Here the index $i$ is used for grouping all the degenerate $g_i$ modes, having the same propagation constant, with photon occupancy $n_i$. Note that, in our MD experiments, we measured the power fraction associated with each mode, which is directly proportional to $n_i$. Therefore, in practice, we consider $n_i$ to be a power fraction, since their proportionality constant will only provide a (superfluous) additional term in Eq. (\ref{eq-S}).

It is evident that the terms in Eq. (\ref{eq-W}) for which $n_i = 0$ do not contribute to the product, since they are equal to 1. Therefore, the entropy can be written as
\begin{equation}
    S = \sum_{i|n_i\neq 0} \ln \frac{(n_i+g_i-1)!}{n_i!(g_i-1)!},
\end{equation}
which, following the same derivation of Ref. \cite{wu2019thermodynamic}, leads to
\begin{equation}
    S = \sum_{i|n_i\neq 0} \ln n_i,
\end{equation}
or equivalently 
\begin{equation}
    S = \sum_{i} S_i,
\end{equation}
where 
\begin{equation}
    S_i = \left\{
\begin{array}{lr}
   \ln n_i & \text{if }  n_i \neq 0\\
    0 & \text{if } n_i = 0 
\end{array}
    \right. .
    \label{eq-Si}
\end{equation}
Experimentally, we often found that some modes have a power fraction close to zero, which makes the logarithmic term in Eq. (\ref{eq-Si}) diverging. For instance, when dealing with thermalized beams, high-order modes are associated with power fractions that are quickly damped when the index $i$ grows larger, as predicted by the RJ distribution. Therefore, in order to meaningfully evaluate the entropy, it is necessary to take into account the tolerances which are imposed by the limited accuracy of our MD. In particular, we call $\Delta n$ the experimental error of our MD method for the estimate of the mode occupancy, and we consider
\begin{equation}
    S_i = \left\{
\begin{array}{lr}
   \ln n_i & \text{if }  n_i \gtrsim \Delta n\\
    0 & \text{if } n_i \lesssim \Delta n 
\end{array}
    \right. .
\end{equation}
Specifically, in the plots of Fig. 4 of the main text, we have set $\Delta n = 0.08$. Whereas the experimental results shown in all other figures take into account the occupancy of all modes. This is because the condition $n_i = 0$ is not detrimental for the accurate estimation of any thermodynamic parameter, except for $S$. 

The values of $T$ and $\mu$ were calculated as described in Ref. \cite{wu2019thermodynamic}: the definition of the total optical power, i.e.,
\begin{equation}
        \mathcal{P} = \sum_i n_i,
\end{equation}
and that of the internal energy, i.e.,
\begin{equation}
       U = \sum_i \beta_i n_i,
\end{equation}
where $\beta_i$ is the propagation constant of the $i$-th mode, are used in the equation of state (i.e., Eq. (1) in the main text) $U-\mu\mathcal{P}=MT$. This provides an equation with two unknowns ($T$ and $\mu$), which, combined with the RJ law, i.e.,
\begin{equation}
    \frac{n_i}{g_i}= -\frac{T}{\mu+\beta_i}.
\end{equation}
leads to a nonlinear system of equations which has an unique solution for $T$ and $\mu$, as long as the physical condition $n_i\geq 0$ is imposed \cite{wu2019thermodynamic}.

\section{Supplementary Note 3:  \\Mode equilibrium distribution}
\begin{figure}[htb]
  \centering
  \includegraphics[width=8.3cm]{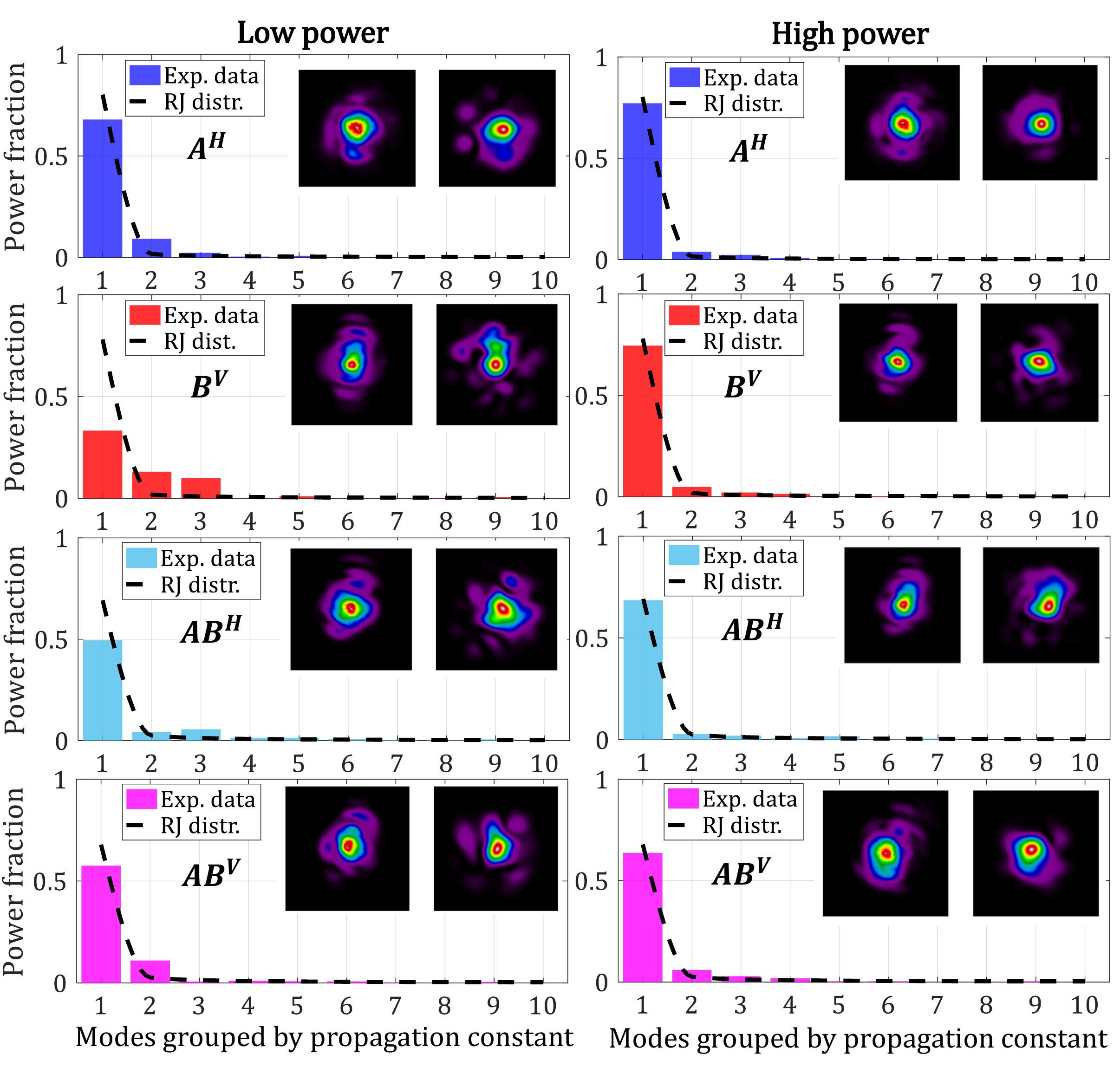}
\caption{MD of the output beam corresponding to the experiments in Fig.3a-c (main text), when operating either in the linear (left column) or in the nonlinear (right column) beam propagation regime. The inset pictures show the measured near-field (left) and their holographic reconstructions (right). The four rows represent different input conditions, i.e., either in presence of the sole A (blue bars) and B (red bars) beams, or when simultaneously injecting both A and B beams (azure and purple bars) into the MMF core, respectively. 
}
  \label{fig-thermo_suppli}
\end{figure}

In Fig.\ref{fig-thermo_suppli}, we report the experimental mode distribution that correspond to the data shown in Fig.3a-c of the main text. For the sake of simplicity, we only show results obtained for the lowest (left column) and for the highest (right column) values of input power, i.e., corresponding to either a nonequilibrium or to a thermal equilibrium state, respectively.
In each sub-figure, we plot as a colored histogram the experimental mode distribution. Whereas the dashed black lines illustrate their associated equilibrium RJ-distribution. 
As it can be seen, at the highest powers the histogram bars are well fitted by the RJ law. To the contrary, at low powers, when thermal equilibrium has not yet been reached, there is a clear discrepancy between the experimental data and the RJ law. We underline that the y axis of the plots in Fig. \ref{fig-thermo_suppli} represents the power fraction in each mode within a group with degeneracy $g_i$. Whereas in our former work (\cite{mangini2022statistical}), we used to plot the total power of each mode group. 
%
Finally, in order to highlight the good quantitative agreement between theory and experiments, we computed the root-mean-square-error (RMSE) between the experimental data and the theoretical RJ distribution. The RMSE values associated with the plots in Fig. \ref{fig-thermo_suppli} are reported in Table \ref{tab1}.
\begin{table}[]
    \centering
    \begin{tabular}{|c|c|c|}
    \hline
      & Lowest power & Highest power\\
    \hline
     $RMSE_{A^H}$ &  4.0 \%  & 3.5 \%\\
    \hline
     $RMSE_{B^V}$ &  8.0 \%  & 3.5 \% \\
    \hline
     $RMSE_{AB^H}$ &  3.4 \% & 2.9 \% \\
    \hline
     $RMSE_{AB^V}$ &  3.8 \% & 3.0 \%\\
    \hline
    \end{tabular}
    \caption{Computed RMSE between the experimental mode decomposition data in Fig.\ref{fig-thermo_suppli} and their associated theoretical RJ curves.}
    \label{tab1}
\end{table}



\end{document}